\documentclass[12pt,prd,tightenlines,nofootinbib,showpacs,showkeys]{revtex4}
\newcommand{\be}{\begin{equation}}
\newcommand{\ee}{\end{equation}}
\usepackage{bm}
\usepackage{graphics}
\usepackage{epsfig}
\begin{document}
\title{\begin{flushright}{\rm\normalsize SSU-HEP-06/02}\end{flushright}
Proton polarizability effect in the hyperfine splitting of the
hydrogen atom}
\author{R.N.Faustov\footnote{faustov@theory.sinp.msu.ru}}
\affiliation{Dorodnicyn Computing Centre RAS, Vavilov Street 40,
Moscow, 119991, Russia}
\author{I.V.Gorbacheva, A. P. Martynenko\footnote{mart@ssu.samara.ru}}
\affiliation{Samara State University, Pavlov Street 1, Samara,
443011, Russia.}

\begin{abstract}
The contribution of the proton polarizability to the ground state
hyperfine splitting in the hydrogen atom is evaluated on the basis
of isobar model and evolution equations for the parton
distributions. The contributions of the Born terms, vector meson
exchanges and nucleon resonances are taken into account in the
construction of the proton polarized structure functions
$g_{1,2}(W,Q^2)$. Numerical values of this effect are equal
$(2.2\pm 0.8)\times 10^{-6}$ times the Fermi splitting in
electronic hydrogen and $(4.70\pm 1.04)\times 10^{-4}$ times the
Fermi splitting in muonic hydrogen.
\end{abstract}



\maketitle

\section{Introduction}

The precise investigation of the energy levels of hydrogenic
atoms (muonium, positronium, hydrogen atom, deuterium, helium
ions et al.) allows to obtain more exact values for many
fundamental physical constants such as the lepton masses, the
ratio of the lepton and proton masses, the fine structure
constant, the Rydberg constant which are used for creating
standards of units \cite{MT}. The insertion of new simple atomic
systems in the range of experimental investigation can lead to
significant progress in solving of these problems. The
measurement of muonic hydrogen Lamb shift at PSI (Paul Sherrer
Institute)  with a precision of 30 ppm will allow to improve our
knowledge of the proton charge radius by an order of the
magnitude \cite{Pohl}. Another important problem is connected
with the measurement of the ground state hyperfine splitting
(HFS) in muonic hydrogen \cite{kp,apm}. In the case of electronic
hydrogen HFS was measured with extremely high accuracy many years
ago \cite{Hellwig}:
\begin{equation}
\Delta E^{exp}_{HFS}=1420~405~751.7667(9)~kHz.
\end{equation}

The corresponding theoretical expression of the hydrogen hyperfine
splitting can be written in the form ($\Delta E^{th}_{HFS}=2\pi\hbar\Delta
\nu^{th}_{HFS}$) \cite{EGS}:
\begin{equation}
\Delta E^{th}_{HFS}=E_F(1+\delta^{QED}+\delta^S+\delta^P),~E_F=\frac{8}{3}
\alpha^4\frac{\mu_Pm_p^2m_e^2}{(m_p+m_e)^3},
\end{equation}
where $\mu_P$ is the proton magnetic moment, $m_e$, $m_p$ are the masses
of the electron and proton. The calculation of different corrections to the
Fermi energy $E_F$ has a long history. Modern status in the theory of
hydrogenic atoms was presented in details in \cite{EGS}. $\delta^{QED}$
denotes the contribution of higher-order quantumelectrodynamical effects.
Corrections $\delta^S$ and $\delta^P$ take into account the influence of
strong interaction. $\delta^S$ describes the effects of proton finite-size
and recoil contribution. $\delta^P$ is the correction due to the proton
polarizability. Basic uncertainties of theoretical result (2) are related
with $\delta^S$ and $\delta^P$.

\begin{figure}[htbp]
\begin{center}
\includegraphics{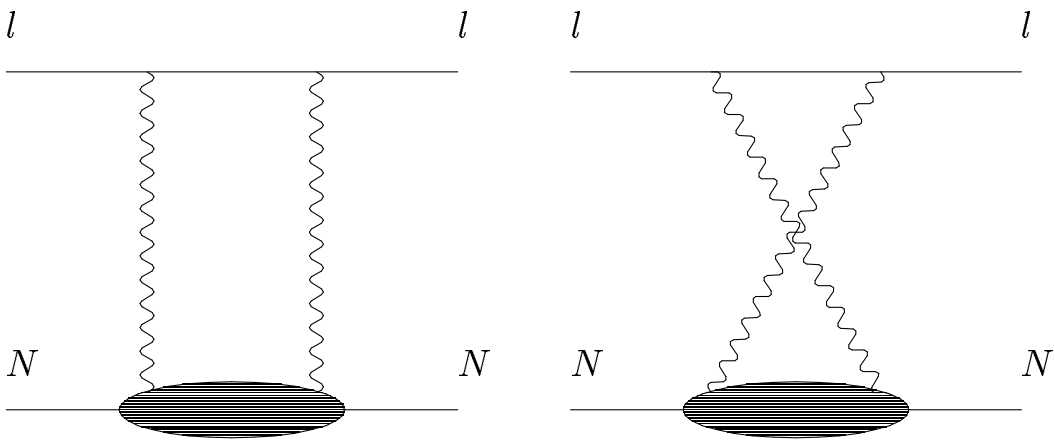}
\end{center}
\bigskip
\caption{Two-photon Feynman amplitudes determining the correction of the
proton polarizability in the hyperfine splitting of the hydrogen atom.}
\end{figure}

The main part of the one-loop proton structure correction is
determined by the following expression (the Zemach correction)
\cite{Zemach}:
\begin{equation}
\Delta E_Z=F_F\frac{2\mu\alpha}{\pi^2}\int\frac{d{\bf p}}{p^4}\left[
\frac{G_E({\bf p}^2)G_M({\bf p}^2)}{1+\kappa}-1\right]=E_F(-2\mu\alpha)R_p.
\end{equation}
The Zemach radius $R_p$ is determined by the densities of
electric charge and magnetic moment. It is considered as a
fundamental parameter of the proton structure along with the
proton charge radius. There exist three possibilities to
determine the numerical value $R_p$
\cite{Friar,Brodsky,Volotka,Bakalov}. One approach based on the
analysis of the world data on $e-p$ scattering gives $R_p =
1.086\pm 0.012$ fm \cite{Friar}. Another method uses the
comparison of theoretical and experimental results (1)-(2) for
the hydrogen atom. In this case the value $R_p=1.043(16)$ fm is
obtained in \cite{Brodsky}. The third method can be based on the
comparison of the future experimental data and theoretical result
for muonic hydrogen \cite{apm,Bakalov,cfm}. The proton
polarizability correction is important among other contributions
of order $10^{-6}$. Numerical estimation of $\delta^P$ obtained
in \cite{MF} serves at present as a reliable guide for defining
the total value of the HFS in the hydrogen
\cite{Friar,Brodsky,Volotka,Bakalov}.

The aim of this work consists in the investigation of the proton
polarizability correction in the hydrogen HFS. We performed new
calculation of $\delta^P$ using the isobar model describing the processes
of photo- and electroproduction of $\pi$, $\eta$ mesons, nucleon resonances
on the nucleon in the resonance region, and on evolution equations for
the parton distributions in deep inelastic region.

\section{General formalism}

The main contribution to $\delta^P$ is determined by two-photon diagrams,
shown in Fig. 1. The corresponding amplitudes of virtual Compton
scattering on the proton can be expressed through nucleon polarized structure
functions $G_1(\nu,Q^2)$ and $G_2(\nu,Q^2)$. Inelastic contribution
of the diagrams (a), (b) Fig. 1 can be presented in the form \cite{MF,SD,V,G,Z}:
\begin{equation}
\Delta E_{HFS}^P=\frac{Z\alpha m_e}{2\pi m_p(1+\kappa)}E_F(\Delta_1+\Delta_2)=
(\delta_1^P+\delta_2^P)E_F=\delta^PE_F,
\end{equation}
\begin{equation}
\Delta_1=\int_0^\infty\frac{dQ^2}{Q^2}\left\{\frac{9}{4}F_2^2(Q^2)-
4m_p^3\int_{\nu_{th}}^\infty\frac{d\nu}{\nu}\beta_1\left(\frac{\nu^2}
{Q^2}\right)G_1(\nu,Q^2)\right\},
\end{equation}
\begin{equation}
\Delta_2=-12m_p^2\int_0^\infty\frac{dQ^2}{Q^2}
\int_{\nu_{th}}^\infty d\nu\beta_2\left(\frac{\nu^2}{Q^2}\right)
G_2(\nu,Q^2),
\end{equation}
where $\nu_{th}$ determines the pion-nucleon threshold:
\begin{equation}
\nu_{th}=m_\pi+\frac{m_\pi^2+Q^2}{2m_p},
\end{equation}
and the functions $\beta_{1,2}$ have the form:
\begin{equation}
\beta_1(\theta)=3\theta-2\theta^2-2(2-\theta)\sqrt{\theta(\theta+1)},
\end{equation}
\begin{equation}
\beta_2(\theta)=1+2\theta-2\sqrt{\theta(\theta+1)},~\theta=\nu^2/Q^2.
\end{equation}
$F_2(Q^2)$ is the Pauli form factor of the proton, $\kappa$ is the proton
anomalous magnetic moment: $\kappa$=1.792847351(28) \cite{MT}.

The polarized structure functions $g_1(\nu,Q^2)$ and $g_2(\nu,Q^2)$ enter
in the antisymmetric part of hadronic tensor ${\rm W_{\mu\nu}}$, describing
lepton-nucleon deep inelastic scattering \cite{Close}:
\begin{equation}
W_{\mu\nu}=W_{\mu\nu}^{[S]}+W_{\mu\nu}^{[A]},
\end{equation}
\begin{equation}
W_{\mu\nu}^{[S]}=\left(-g_{\mu\nu}+\frac{q_\mu q_\nu}{q^2}\right)W_1(\nu,Q^2)+
\left(P_\mu-\frac{P\cdot q}{q^2}q_\mu\right)\left(P_\nu-\frac{P\cdot q}{q^2}q_\nu\right)
\frac{W_2(\nu,Q^2)}{m_p^2},
\end{equation}
\begin{equation}
W_{\mu\nu}^{[A]}=\epsilon_{\mu\nu\alpha\beta}q^\alpha\left\{S^\beta
\frac{g_1(\nu,Q^2)}{P\cdot q}+[(P\cdot q)S^\beta-(S\cdot q)P^\beta]
\frac{g_2(\nu,Q^2)}{(P\cdot q)^2}\right\},
\end{equation}
where $\epsilon_{\mu\nu\alpha\beta}$ is the totally antisymmetric
tensor in four dimensions, $g_1(\nu,Q^2)=m_p^2\nu G_1(\nu,Q^2)$,
$g_2(\nu,Q^2)=m_p\nu^2 G_2(\nu,Q^2)$, P is the four-momentum of the
nucleon, $x=Q^2/2m_p\nu$ is the Bjorken variable, $S$ is the proton
spin four-vector, normalized to $S^2=-1$, $q^2=-Q^2$ is the square
of the four-momentum transfer. The invariant quantity $P\cdot q$ is
related to the energy transfer $\nu$ in the proton rest frame:
$P\cdot q=m_p\nu$. The invariant mass of the electroproduced
hadronic system, W, is then
$W^2=m_p^2+2m_p\nu-Q^2=m_p^2+Q^2(1/x-1)$. Here $W_1$ and $W_2$ are
the structure functions for unpolarized scattering. In the DIS
regime the invariant mass W must be greater than any resonance in
the nucleon. The threshold between the resonance region and the
deep-inelastic region is not well defined, but it is usually taken
to be at about $W^2=4~GeV^2$.

Hadronic tensor $W_{\mu\nu}$ is proportional to the
imaginary part of the off-shell Compton amplitude for the forward
scattering of virtual photons on nucleons: $\gamma^\ast N\rightarrow
\gamma^\ast N$. The photon-nucleon interaction
depends on the photon polarization as well as on the nucleon one.
This gives four independent helicity amplitudes of the
form $M_{ab;cd}$, with a, b, c, d values for the helicities
of the photon and nucleon initial and final states:
\begin{displaymath}
M_{1,1/2;1,1/2},~~~M_{1,-1/2;1,-1/2},~~~M_{0,1/2;0,1/2},~~~M_{1,1/2;0,-1/2}.
\end{displaymath}
These components correspond to the four structure
functions $W_1, W_2, g_1, g_2$. All other possible
combinations of initial and final photon and nucleon helicities
are related to the above by time reversal and parity transformation.

The proton spin structure functions can be measured in the
inelastic scattering of polarized electrons on polarized protons.
Recent improvements in polarized lepton beams and nucleon targets
have made it possible to make accurate measurements of nucleon
polarized structure functions $g_{1,2}$ in experiments at SLAC,
CERN and DESY \cite{Abe1,Abe2,Anthony,Mitchell,Adams,Adeva}. The
spin dependent structure functions can be expressed in terms of
virtual photon-absorption cross sections \cite{Close}:
\begin{equation}
g_1(\nu,Q^2)=\frac{m_2\cdot K}{8\pi^2\alpha(1+Q^2/\nu^2)}\left[\sigma_{1/2}
(\nu,Q^2)-\sigma_{3/2}(\nu,Q^2)+\frac{2\sqrt{Q^2}}{\nu}\sigma_{TL}(\nu,Q^2)\right],
\end{equation}
\begin{equation}
g_2(\nu,Q^2)=\frac{m_2\cdot K}{8\pi^2\alpha(1+Q^2/\nu^2)}\left[-\sigma_{1/2}
(\nu,Q^2)+\sigma_{3/2}(\nu,Q^2)+\frac{2\nu}{\sqrt{Q^2}}\sigma_{TL}(\nu,Q^2)\right],
\end{equation}
where $K$ is the kinematical flux
factor for virtual photons, $\sigma_{1/2}$, $\sigma_{3/2}$ are
the virtual photoabsorption transverse cross sections for the total
photon-nucleon helicity of 1/2 and 3/2 respectively,
$\sigma_{TL}$ is the interference term between the
transverse and longitudinal photon-nucleon amplitudes. In this
work we calculate contribution $\Delta E_{HFS}^P$ on the
basis of the latest experimental data on the structure functions $
g_{1,2}(\nu,Q^2)$ and theoretical predictions for the cross sections
$\sigma_{1/2,3/2,TL}$.

The proton polarizability contribution to HFS in the resonance
region is determined by the processes of photo- and
electroproduction on nucleons of the pions and some prominent
baryon resonances. The amplitudes of such reactions are shown in
Fig.2.

\begin{figure}[htbp]
\begin{center}
\includegraphics{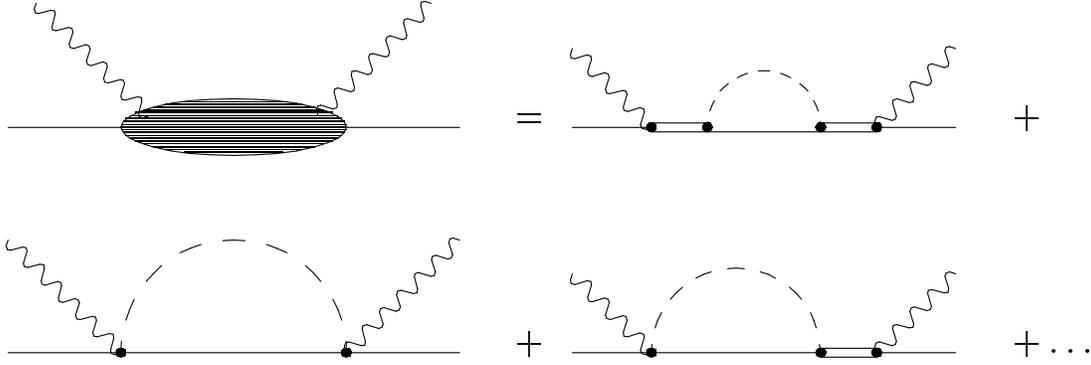}
\end{center}
\bigskip
\caption{Feynman amplitudes for the proton
polarizability correction in the resonance region. Solid, double
solid, wave and dashed lines correspond to nucleon, baryon
resonance, photon and pion correspondingly.}
\end{figure}

To obtain correction (4) in the resonance region $(W^2\leq 4
GeV^2)$ we use the Breit-Wigner parameterization for the
photoabsorption cross sections in Eqs.(13)-(14), suggested in
\cite{Walker,Arndt,Teis1,Teis2,Krusche,Bianchi,D,Dong}. In the
considered region of the variables $k^2$, $W$ the most
contribution is given by five resonances: $P_{33} (1232)$,
$S_{11} (1535)$, $D_{13} (1520)$, $P_{11} (1440)$, $F_{15}
(1680)$. Accounting the resonance decays to the $N\pi-$ and
$N\eta-$ states we can express the absorption cross sections
$\sigma_{1/2}$ and $\sigma_{3/2}$ as follows:
\begin{equation}
\sigma_{1/2,3/2}=\left(\frac{k_R}{k}\right)^2\frac{W^2\Gamma_\gamma\Gamma_{R
\rightarrow N\pi}}{(W^2-M_R^2)^2+W^2\Gamma_{tot}^2}\frac{4m_p}{M_R\Gamma_R}
|A_{1/2,3/2}|^2,
\end{equation}
where $A_{1/2,3/2}$ are transverse electromagnetic helicity amplitudes,
\begin{equation}
\Gamma_\gamma=\Gamma_R\left(\frac{k}{k_R}\right)^{j_1}\left(\frac{k_R^2+X^2}
{k^2+X^2}\right)^{j_2},~~X=0.3~GeV.
\end{equation}
The resonance parameters $\Gamma_R$, $M_R$, $j_1$, $j_2$,
$\Gamma_{tot}$ are taken from \cite{PDG,Teis1,Teis2,Teis3}. In
accordance with \cite{Teis1,Krusche,Teis3} the parameterization
of one-pion decay width is
\begin{equation}
\Gamma_{R\rightarrow N\pi}(q)=\Gamma_R\frac{M_R}{M}\left(\frac{q}{q_R}\right)^3
\left(\frac{q_R^2+C^2}{q^2+C^2}\right)^2,~C=0.3~GeV
\end{equation}
for the $P_{33}(1232)$ and
\begin{equation}
\Gamma_{R\rightarrow N\pi}(q)=\Gamma_R\left(\frac{q}{q_R}\right)^{2l+1}
\left(\frac{q_R^2+\delta^2}{q^2+\delta^2}\right)^{l+1},
\end{equation}
for $D_{13}(1520)$, $P_{11}(1440)$, $F_{15}(1680)$. $l$ is the pion angular
momentum and $\delta^2$ = $(M_R-$ $m_p-m_\pi)^2$ + $\Gamma_R^2/4$. Here $q$
$(k)$ and $q_R$ $(k_R)$ denote the c.m.s. pion (photon) momenta of resonances
with the mass M and $M_R$ respectively. In the case of $S_{11}(1535)$ we take
into account $\pi N$ and $\eta N$ decay modes \cite{Krusche,Teis3}:
\begin{equation}
\Gamma_{R\rightarrow\pi,\eta}=\frac{q_{\pi,\eta}}{q}b_{\pi,\eta}\Gamma_R
\frac{q_{\pi\eta}^2+C_{\pi,\eta}^2}{q^2+C_{\pi,\eta}^2},
\end{equation}
where $b_{\pi,\eta}$ is the $\pi$ ($\eta$) branching ratio.

The cross section $\sigma_{TL}$ is determined by an expression
similar to Eq.(15), containing the product $(S^\ast_{1/2}\cdot
A_{1/2}+A_{1/2}^\ast S_{1/2})$ \cite{Abe1}. The calculation of
helicity amplitudes $A_{1/2}$, $A_{3/2}$ and longitudinal
amplitude $S_{1/2}$, as functions of $Q^2$, was done on the basis
of constituent quark model (CQM) in
\cite{Dong2,Isgur,CL,Capstick,LBL,Warns}. In the real photon
limit $Q^2=0$ we take corresponding resonance amplitudes from
\cite{PDG}.

\begin{figure}[htbp]
\begin{center}
\includegraphics{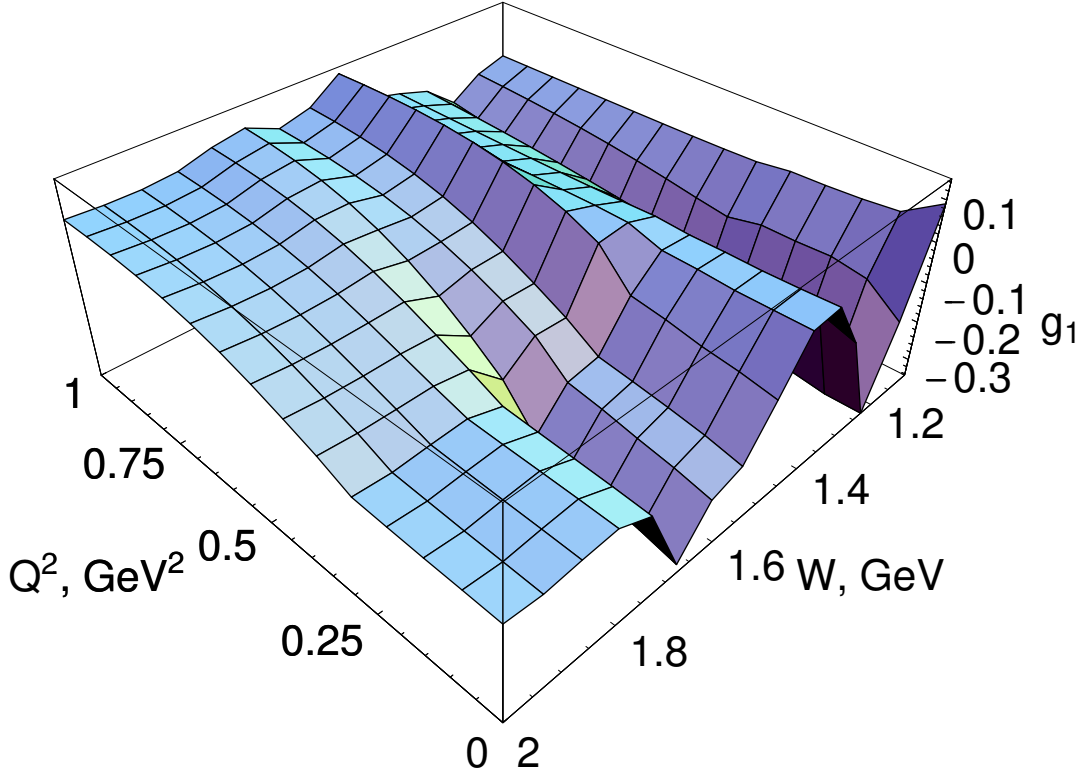}
\end{center}
\caption{The proton polarized structure function $g_1(W,Q^2)$ as the
function of variables $Q^2$ ($0\div 1$) $Gev^2$ and $W$ ($1.1\div
2.0$) $GeV$.}
\end{figure}

\begin{figure}[htbp]
\begin{center}
\includegraphics{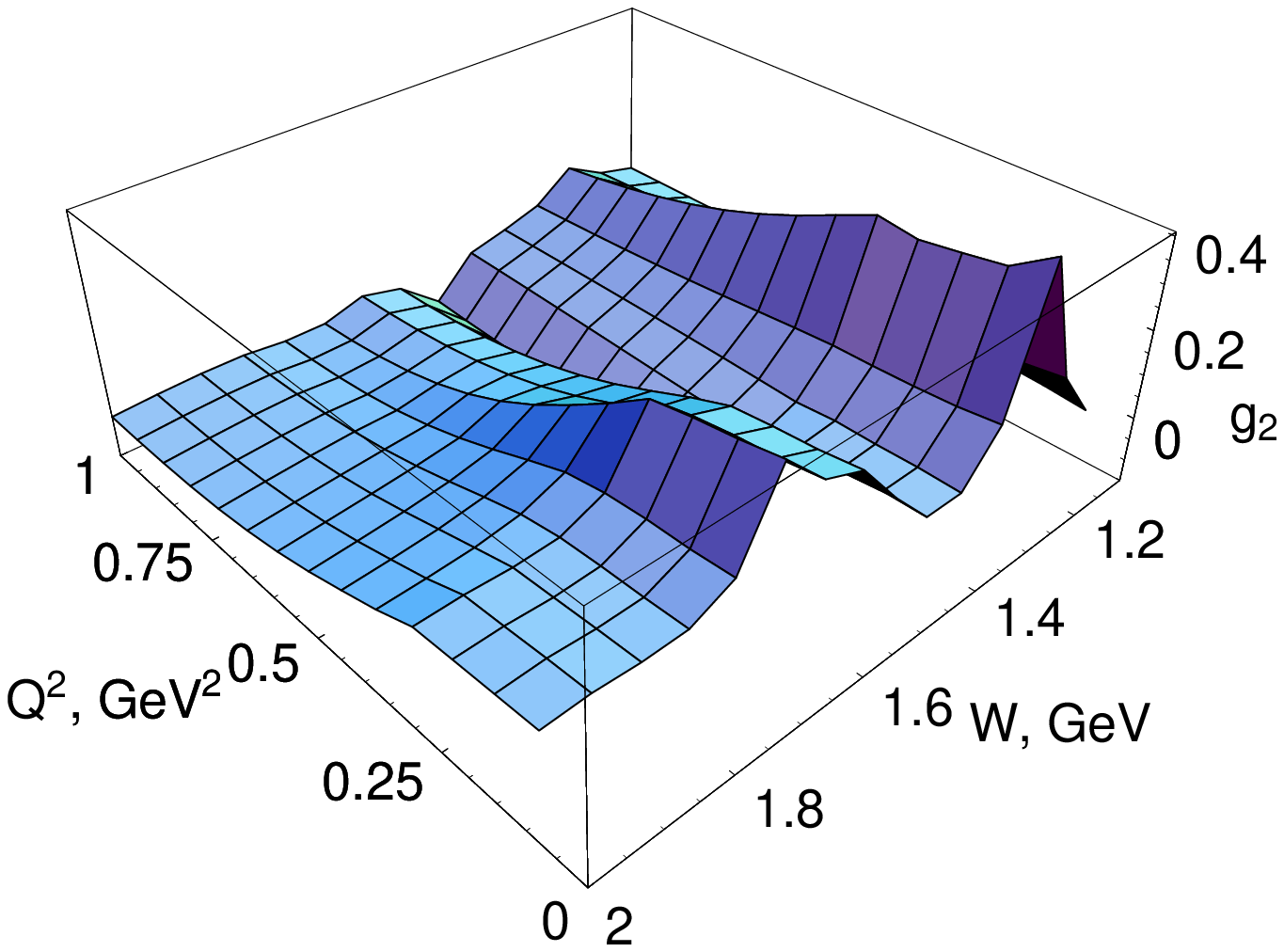}
\end{center}
\caption{The proton polarized structure function $g_2(W,Q^2)$ as the
function of variables $Q^2$ ($0\div 1$) $Gev^2$ and $W$ ($1.1\div
2.0$) $GeV$.}
\end{figure}

The two-pion decay modes of the higher nucleon resonances
($S_{11}(1535)$, $D_{13}(1520)$, $P_{11}(1440)$ and $F_{15}(1680)$
were described phenomenologically using two-step process as in
\cite{Teis1}. The high-lying nucleon resonance $R$ can decay first
into $N^{\ast}$ ($ P_{33}(1232)$ or $P_{11}(1440)$) and a pion or
into a nucleon and $\rho$- or $\sigma$-meson. Then the new
resonances decay into a nucleon and a pion or two pions:
\begin{displaymath}
R\rightarrow r+a=\Biggl\{{N^\ast+\pi\rightarrow
N+\pi+\pi,\atop \rho(\sigma)+N\rightarrow N+\pi+\pi.}
\end{displaymath}
The total decay width of such processes can be presented as a
phase space weight integral over the mass distribution of the
intermediate resonance $r$ = $N^\ast, \rho, \sigma$ ($a=\pi, N$):
\begin{equation}
\Gamma_{R\rightarrow r+a}(W)=\frac{P_{2\pi}}{W}\int_0^{W-m_a}d\mu \cdot
p_f\frac{2}{\pi}
\frac{\mu^2\Gamma_{r,tot}(\mu)}{(\mu^2-m_r^2)^2+\mu^2\Gamma_{r,tot}^2(\mu)}
\frac{(M_R-m_2-2m_\pi)^2+C^2}{(W-m_2-2m_\pi)^2+C^2},
\end{equation}
where $C=0.3~GeV$, the factor $P_{2\pi}$ must be taken from the constraint
condition: $\Gamma_{R\rightarrow r+a}(W_R)$ coincides with the experimental
data, $p_f$ is the three momentum of the resonance $r$ in the rest frame of
$R$. $\Gamma_{r,tot}$ is the total width of the resonance $r$. The
decay width of the meson resonance is parameterized similarly to
that of the $P_{33}(1232)$:
\begin{equation}
\Gamma(\mu)=\Gamma_r\frac{m_r}{\mu}\left(\frac{q}{q_r}\right)^{2J_r+1}
\frac{q_r^2+\delta^2}{q^2+\delta^2},~~~\delta=0.3~GeV,
\end{equation}
where $m_r$ and $\mu$ are the mean mass and the actual mass of the meson
resonance, $q$ and $q_r$ are the pion three momenta in the rest frame of the
resonance with masses $\mu$ and $m_r$, $J_r$ and $\Gamma_r$
are the spin and decay width of the resonance with the mass $m_r$.

Main nonresonant contribution to the cross sections $\sigma_{T,L}$
in the resonance region is determined by the Born terms
constructed on the basis of Lagrangians of $\gamma NN$,
$\gamma\pi\pi$, $\pi NN$ interactions. Another part of
nonresonant background comprises the $t$- channel contributions
of $\rho$, $\omega$ mesons obtained by means of effective
Lagrangians $\gamma \pi V$, $VNN$ interactions ($V=\rho,\omega$)
\cite{D}. In the unitary isobar model accounting the Born terms,
the vector meson, nucleon resonance contributions and the
interference terms we calculated the cross sections
$\sigma_{T,L}$ by means of numerical program MAID
(http://www.kph-uni-mainz.de/MAID) in the resonance region as the
functions of two variables $W$ and $Q^2$. The obtained nucleon
polarized structure functions $g_{1,2}(W,Q^2)$ are presented in
Figs.3-4. These results for the structure function $g_1(W,Q^2)$
are in qualitative agreement with experimental data. The
particular significance in the study of the spin-dependent
properties of baryon resonances belongs to Gerasimov-Drell-Hearn
(GDH) sum rule \cite{GDH}
\begin{equation}
-\frac{\kappa^2}{4m_2^2}=\frac{1}{8\pi^2\alpha}\int_{\nu_{th}}^\infty
\frac{d\nu}{\nu}[\sigma_{1/2}(\nu,0)-\sigma_{3/2}(\nu,0)].
\end{equation}
The GDH sum rule rests on the basic physical principles and an unsubtracted
dispersion relation applied to the forward Compton amplitude.

To construct the spin-dependent structure functions of the proton in deep
inelastic region we can use the $Q^2$ evolution equations for the quark
and gluon distributions \cite{DGLAP}:
\begin{equation}
\frac{dq_i(x,Q^2)}{d\ln Q^2}=\frac{\alpha_s}{2\pi}\int_x^1\frac{dy}{y}
\left[q_i(y,Q^2)P_{qq}\left(\frac{x}{y}\right)+g(y,Q^2)P_{qg}\left(
\frac{x}{y}\right)\right],
\end{equation}
\begin{equation}
\frac{dg(x,Q^2)}{d\ln Q^2}=\frac{\alpha_s}{2\pi}\int_x^1\frac{dy}{y}
\left[\sum_iq_i(y,Q^2)P_{gq}\left(\frac{x}{y}\right)+g(y,Q^2)P_{gg}\left(
\frac{x}{y}\right)\right],
\end{equation}
where the sum is considered over all quarks and antiquarks.
$P_{qq}$, $P_{gq}$, $P_{qg}$, $P_{gg}$ are the quark-gluon
splitting functions \cite{LP}. Numerical solution of the
integrodifferential evolution equations (23), (24) by means of the
method suggested in \cite{Kumano} allows to obtain the parton
distributions and the structure functions $g_{1,2}(x,Q^2)$ for
different values of a photon momentum squared $Q^2$.
Corresponding numerical results are in good agreement with the
world experimental data
\cite{Abe1,Abe2,Anthony,Mitchell,Adams,Adeva}.

\begin{figure}[htbp]
\centerline{\psfig{figure=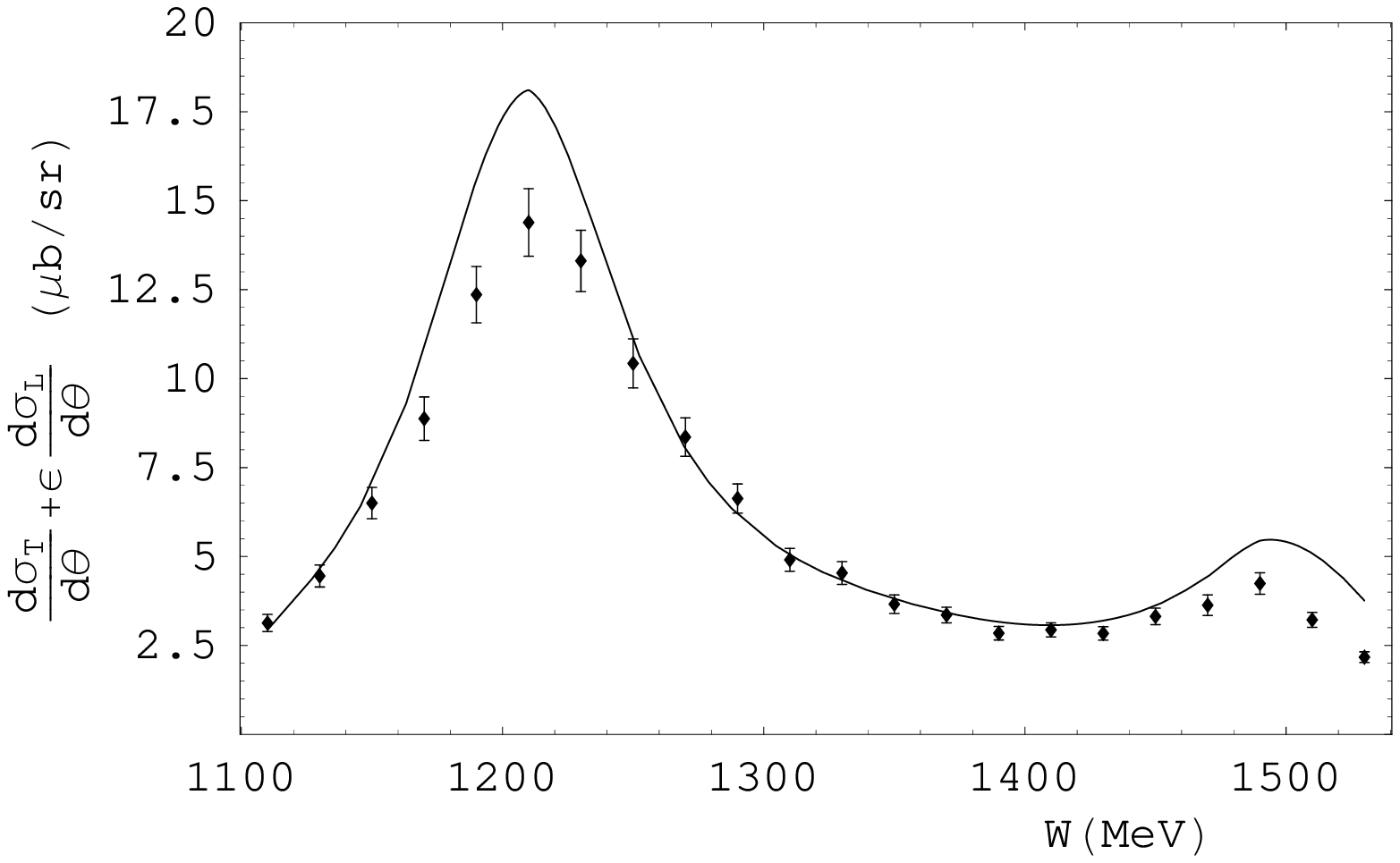,height=6.5cm,width=11.0cm}}
%\centerline{\includegraphics[height=6cm,width=8.5cm]{SigW.eps}}
\hspace{5mm}
\centerline{\psfig{figure=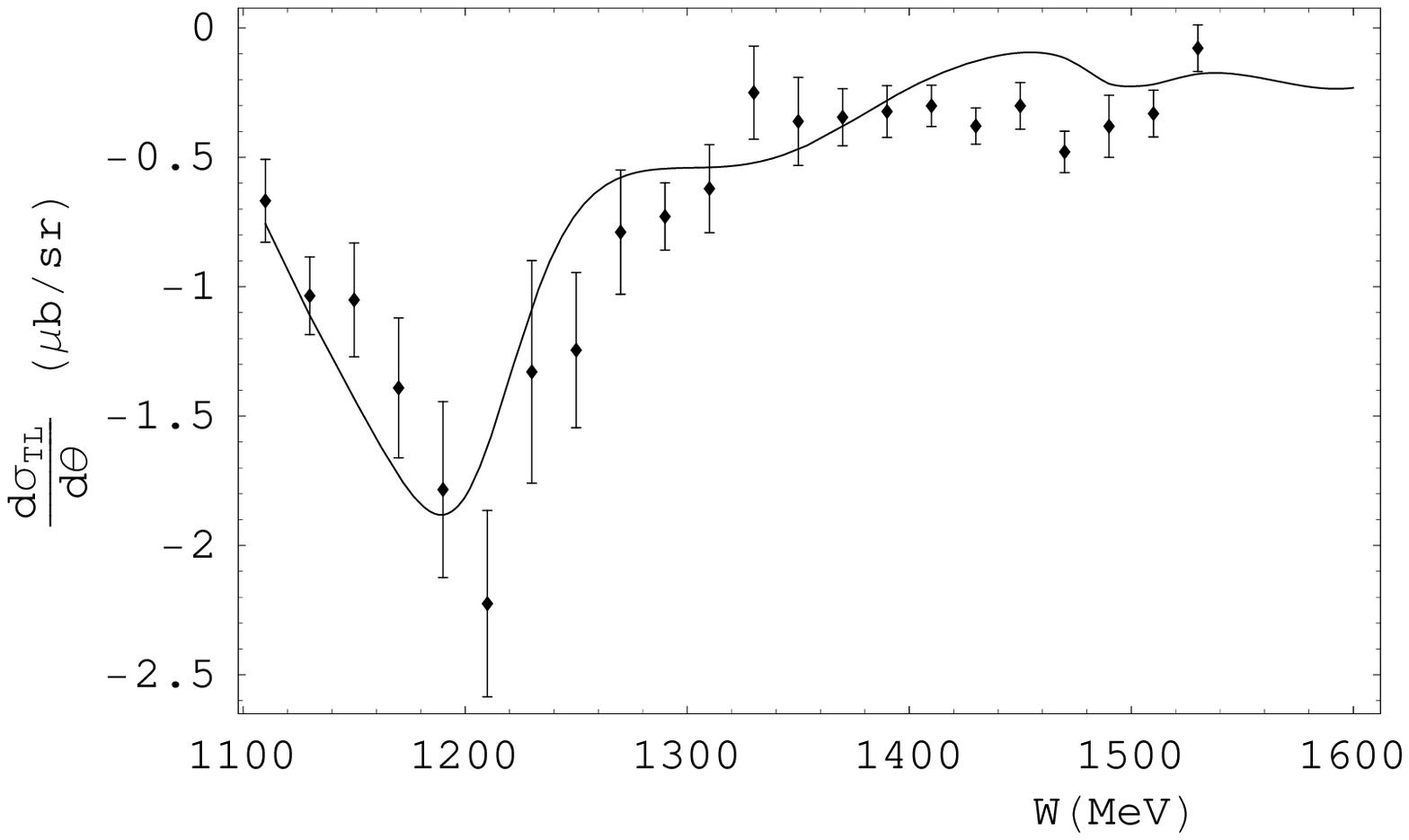,height=6.5cm,width=11.0cm}}
%\centerline{\includegraphics[height=6cm,width=8.5cm]{SigTL.eps}}
\hspace{5mm}
\centerline{\psfig{figure=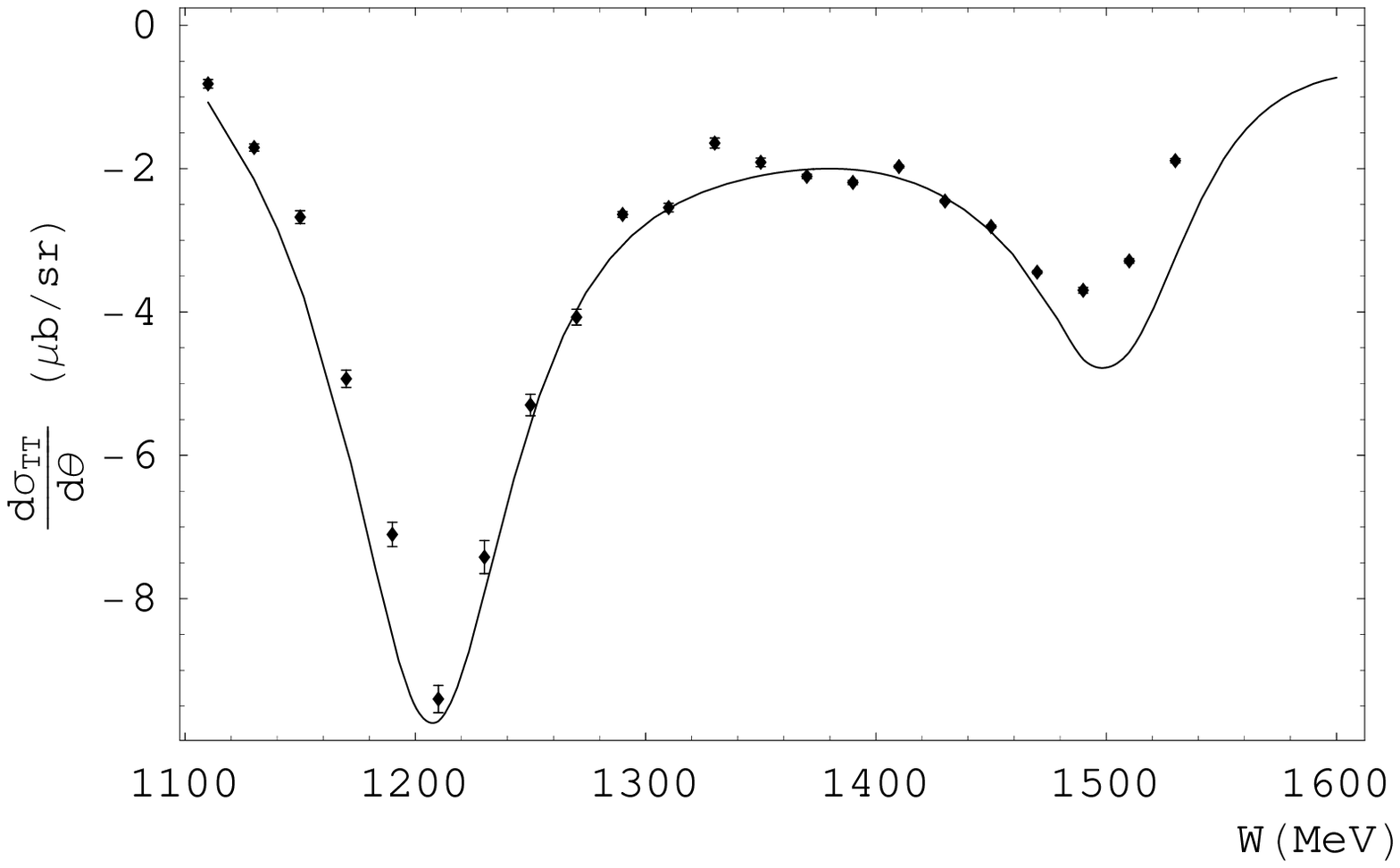,height=6.5cm,width=11.0cm}}
%\centerline{\includegraphics[height=6cm,width=8.5cm]{SigTT.eps}}
\hspace{3mm} \caption[]{\label{fg:CLAS} Plots of $\pi^{+}$ virtual
photoproduction cross sections at $Q^2=0.4$ $GeV$ as the functions
of the $W$ ($1100\div 1600$) $MeV$, obtained on CLAS detector. The
solid curves correspond to the results from MAID.}
\end{figure}

\section{Numerical results}

In this paper we calculate the proton polarizability correction
to the hyperfine splitting of the ground state in the hydrogen
atom on the basis of the isobar model describing the processes of
low-energy scattering of virtual photons on nucleons and the
evolution equations for the parton distributions. These two
significant ingredients of the calculation allow to construct the
absorption cross sections for transversely and longitudinally
polarized photons by nucleons and to express the structure
functions $g_{1,2}(W,Q^2)$ (13), (14), which determine required
contribution (4).

The values of contributions $\delta_1^P$, $\delta_2^P$ and the
total contribution $\delta^P$, obtained after the numerical
integration in the resonance and nonresonant regions are as
follows in electronic and muonic hydrogen correspondingly:
\begin{equation}
\delta_{1}^P=2.6~ppm,~~\delta_{2}^P= -0.4~ppm,~~\delta^P=(2.2\pm
0.8)~ppm,
\end{equation}
\begin{equation}
\delta_{1}^P=5.18\times 10^{-4},~~\delta_{2}^P= -0.48\times
10^{-4},~~\delta^P=(4.70\pm 1.04)\times 10^{-4}.
\end{equation}

The difference of obtained number for electronic hydrogen from the
result of our previous work $(1.4\div 0.6)~ ppm$ \cite{MF} is
connected with new contributions (nonresonant background in the
resonance region and two pion decays of the resonances) which we
considered in this study. There exists a number of theoretical
uncertainties associated with quantities entering in the
correction (4). In the improved isobar model \cite{D,UIM2,UIM3}
containing 14 resonances, we can omit theoretical error which
arises due to the insertion of other high-lying nucleon
resonances. On our sight the main theoretical error is closely
related with the calculation of the helicity amplitudes
$A_{1/2}(Q^2)$, $A_{3/2}(Q^2)$, $S_{1/2}(Q^2)$ in the quark model
based on the oscillator potential \cite{Close}. Only systematical
experimental data for the helicity amplitudes of the
photoproduction on the nucleons $A_{1/2}(0)$, $A_{3/2}(0)$ are
known with sufficiently high accuracy to the present \cite{PDG}.
In the case of amplitudes for the electroproduction of the nucleon
resonances experimental data contain only their values at several
points $Q^2$. So, we have no consistent check for the predictions
of the oscillator model. Possible theoretical uncertainty
connected with the calculation of amplitudes $A_{1/2}(Q^2)$,
$A_{3/2}(Q^2)$, $S_{1/2}(Q^2)$ with the account of relativistic
corrections can attain the value of order $10\%$. Then the
theoretical error for the correction (4) in the resonance region
comprises $20\%$ from the obtained value. We solved DGLAP
equations in the NLO approximation, so possible uncertainty in
$\delta^P$ can comprise near 10$\%$ of obtained result in the
nonresonant region. The other source of the theoretical
uncertainty arises from the experimental data errors in the $
Q^2\le 1~GeV^2$ region. We estimated it at a level of about $20\%$
of the contribution $\delta^P$ at $Q^2\le 1~GeV^2$ in the
nonresonant region.

New experimental data for the electroproduction cross sections in
the reaction $ep\to e'\pi^+n$ in the resonance region were obtained
recently on CLAS detector \cite{CLAS}. Their comparison with the
calculations carried out on the basis of the unitary isobar model
(MAID) is presented in Fig.5. It evidently shows that MAID, which we
used in our calculation, gives the numerical values for the
electroproduction cross sections in the regions of the $\Delta$
isobar and resonances $D_{11}(1520)$, $S_{11}(1535)$, which are
slightly higher (approximately by $5\%$) than the experimental data.

The work was supported by the Russian Foundation for Basic Research
(grant No. 06-02-16821).

\end{document}